# Using Functional Programming for Development of Distributed, Cloud and Web Applications in F#

An overview of functional programming beauty of the web

Dmitri Soshnikov
Developer and Platform Evangelism Unit, Microsoft
Department of Software Engineering, Higher School of Economics
Moscow, Russia
dmitri@soshnikov.com

*Abstract in English*— In this paper, we argue that modern functional programming languages – in particular, F# on the .NET platform – are well suited for the development of distributed, web and cloud applications on the Internet. We emphasize that F# can be successfully used in a range of scenarios – starting from simple ASP.NET web applications, and including cloud data processing tasks and data-driven web applications. In particular, we show how some of the F# features (eg. quotations) can be effectively used to develop a distributed web system using single code-base, and describe the commercial WebSharper project by Intellifactory for building distributed client-server web applications, as well as research library that uses Windows Azure for parametric sweep computational tasks.

*Keywords – distributed systems;web programming; functional programming; f#;javascript;quotations; websharper; windows azure; cloud computing*

*Abstract in Russian*— В данной статье мы хотим привлечь внимание читателя к тому факту, что современные функциональные языки программирования – в частности F# на платформе Microsoft .NET – могут быть успешно применены для разработки распределенных, облачных и веб-приложений. Мы рассмотрим набор типовых сценариев – начиная от простого веб-приложения на ASP.NET, и заканчивая задачами облачной обработки данных и созданием приложений, ориентированных на данные. Мы покажем, как некоторые возможности F# (в частности, квотирование), могут быть эффективно использованы для построения распределенных приложений с использованием единой базы кода, и как эта идея развивается в проекте WebSharper для построения распределенных клиент-серверных веб-систем. Также мы упомянем исследовательский проект библиотеки распределенных вычислений в Windows Azure для задач класса parametric sweep.

*Keywords – распределенные приложения; веб-программирование; функциональное программирование; F#;websharper; квотирование*

## I. Introduction

Most of the complex software systems today are inherently distributed: be that simple distribution between rich client and server located somewhere in the cloud, or more complex interoperability between processing nodes in the cluster. Development of distributed systems is very complex in nature, partly due to the fact that different languages and technologies are used on client and server side, and often there is no transparent seamless integration between those technologies. Thus developer has to think of the whole system as a number of interoperating blocks, always keeping in mind how those block work together and exchange information. This also makes development more complex, because we need to separate server-side and client-side code in the source tree.

To overcome those problems, one of the proposed solutions is to create a brand new programming language, like recently announced Dart [1], which can be used both as server-side language via traditional virtual machine byte-code approach, or on the client via translation to ECMAScript.

However, creating a new programming language is not a necessity. The problem of pragmatic usage of distributed programming languages and approaches has been studied in [2], where it is shown that the problem can be handled on several abstraction levels, from low-level TCP/IP stack to high-level distribution constructs of Distributed Parallel Haskell.

In this paper, we consider the use of F# in the development of distributed web systems. We show how F# can be used to organize distributed code execution on server and client side, giving as an example commercial WebSharper project [3]. We believe that developers and system architects alike would benefit from realizing the potential and beauty of F# language as unified language for the distributed web.

## II. Distribution patterns and three screens principle

Complex software systems nowadays typically exhibit rich user interfaces available on multiple platforms and devices, coupled with massive processing power and storage capacity of

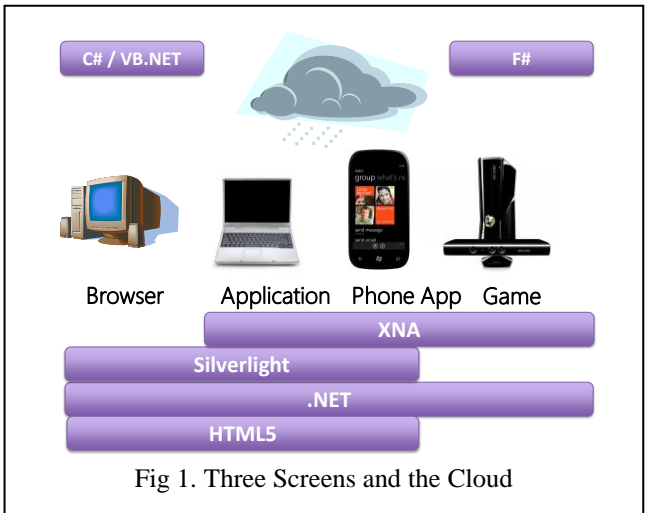

Fig 1. Three Screens and the Cloud

the web and the cloud. This view of the world is known as "three screens and the cloud", demonstrated on Fig. 1

When considering Microsoft platform for those scenarios, we can see that all devices can be programmed using .NET platform, which make F# potentially very useful language. We will argue in this paper that F# opens additional advantages over traditional .NET languages such as C#, in particular related to developing cross client-server functionality.

In three screens paradigm, we can observe several distribution points in complex client-server system:

- Distribution between client code running on .NET programmable device and the server side. In this scenario, traditional web services are a good solution, and it can be easily implemented in F#. One particularly important case of this distribution is simple web site (or cloud web site), which produces HTML web page as a result of a call. This is considered in section III.

- Distribution between web server data processing code and rich AJAX-based web client. In this case JSON asynchronous communication is commonly used, and data has to be passed between C# code and JavaScript code on the client in non-transparent manner. This scenario can be handled using WebSharper that allows developing the whole system in F# using high-level abstractions, with client parts automatically translated to JavaScript. WebSharper is briefly described in section V.

- Distribution between several agents / processes running in parallel in the cloud. We will give the brief example of the project where F# was used to parallelize the computational task and distribute it over several computational Windows Azure working roles. An example of cloud-based distributed scientific computing is given in section VI.

In most of those cases, meta-programing features of F# can be greatly used, in particular quotations. The role of quotations for code distribution is considered in section IV.

It has to be noted, that F# is also a suitable language for programming devices (Windows Phone 7, XNA Grahic Clients for Windows and XBox, etc.). However, we will not consider using F# for devices in this paper – it is described in more detail in [5].

### III. SIMPLE WEB PROGRAMMING IN F#: ASP.NET WEB FORMS AND MVC

Functional programming is in general a good choice for web development, because it is very similar in nature to the stateless functional programming paradigm. A function computes the result given its input data; in a similar manner stateless web page has to produce itself given the input parameters. Such functional web frameworks exist in several functional languages (for example of using LISP/Scheme for web development, see [4]). However, F# is more traditionally used with more common Microsoft web stack technologies, such as ASP.NET Web Forms and ASP.NET MVC. We will not concentrate on those scenarios, as they are pretty straightforward and well described in [5]. It has to be noted that ASP.NET MVC can be used particularly well with F#, since models and controllers (that constitute large part of system's logic) are essentially data processing components that are well developed in functional language. This can also be said about unit testing, which is often used with larger scale ASP.NET MVC projects.

In most of those cases, meta-programing features of F# can be greatly used, in particular quotations.

### IV. QUOTATIONS AND CODE DISTRIBUTION

One of the great meta-programming features of F# are quotations [5,6], which are essentially a way to keep F# program code tree intact in uncompiled form, left for further processing during runtime. Quotations are syntactically separated by <@ and @> symbols, and such piece of code returns quotation of the type Expr<'t> (where 't is the type of code inside the quotation).

Quotation tree can be then traversed to produce the code in a different language, or manipulated in more complex ways to modify code at runtime. When needed, quotation can be compiled and executed using special quotation evaluator available as part of LINQ in the F# Powerpack project [7]. Different ways in which quotations can be used in such a manner are presented in [8], including examples of distributed execution.

As an example, we consider RQXS (Remote Quotation Execution Service) project for simple quotation remote execution has been developed by the author and his students at MAILabs laboratory. It provides a library that can be used to execute quoted pieces of F# code at the dedicated web server and/or cloud cluster. It serializes the quotation, schedules the node for its execution, sends it to that node, where the quotation is deserialized and evaluated. For example, the following code performs the computation on the remote server:

```
let ex = new РExecutor<int>()
ex.Eval(<@ 1+2+3 @>) |> printfn "%A"
ex.Eval <@ [1..100] |> List.sum @> |> printfn "%A"
[1;2;3;4]
   |> List.map (fun x -> ex.Eval <@ x+1 @>)
   |> printfn "%A"
```

Here RExecutor encapsulates the remote computing object of the given type, which can be used to evaluate the quotation remotely. Note that in the last example quotation evaluation is embedded into a series of computations. In the similar manner it can be used to schedule execution of a number of requests (however, to do this properly, we need to switch to asynchronous execution, which is well supported by F# asynchronous workflow syntax).

Even though the examples above are very simple, more complex code can be embedded in the quotation, as in example below:

```
let ex = new RExecutor<int>()
[1..30]
 |> List.map
     (fun z ->
        ex.Eval
          <@ let rec fib x =
              if x<2 then 1
              else fib (x-1)+fib(x-2)
            in fib z @>)
 |> printfn "%A"
```

At the present stage of development, described project is more of a proof of concept than the complete library. In the future it provides good directions for further research and experimentation. For example, in the future the execution service can be extended to include the graphics co-processors in the manner described in [8], or the remote execution of code in-browser via WebSharper. Below is the architectural diagram of the system using RQXS project:

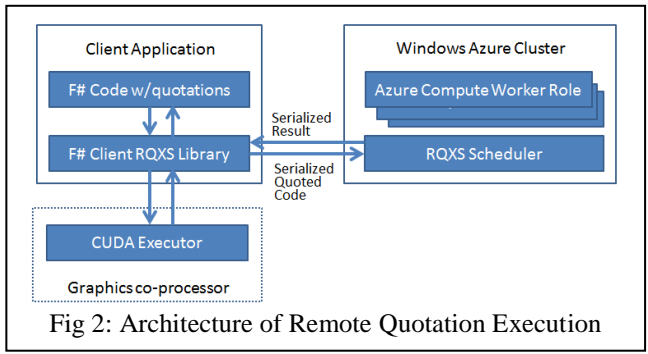

Fig 2: Architecture of Remote Quotation Execution

## V. USING QUOTATIONS FOR WEB CLIENT PROGRAMMING: WEBSHARPER

One of the very important classes of distributed systems are complex rich web applications. In such applications, some functionality is handled by client-side code (typically written in Javascript), while most of processing is done on the server side. Using JavaScript has some disadvantages, namely:

- Using different language for the client side does not allow transferring or re-using parts of the code
- JavaScript lacks brevity and rich type system of F#, and does not allow building rich functional abstractions

Websharper project by Intellifactory [3] uses the following major ideas to unify in one language the page layout (HTML), client code (JavaScript) and server-side code (F#):

- quotations are used to translate parts of F# code into Javascript to be executed in the browser
- transparent function calls are allowed between client and server-side code
- domain-specific language resembling HTML is used for page layout. On top of this language additional functional abstractions can be built to describe larger page blocks and behaviours.

One of the main reasons we describe the Websharper project here is to demonstrate how beautifully different features of F# can be used in order to make this language a unified language for web applications. We believe that knowing the power of Webshaper would allow developers and architects to consider it as a basis for their projects, which would eventually make development more error-prone and faster.

### A. Domain-Specific Language for page layout

Let us see the example of simplest WebSharper project. The main page of the application is built using traditional ASP.NET manner with WebSharper control responsible for rendering the main part of the page (while ASP.NET can handle the major parts of the design). The main page is described in F# in the following manner:

```
[<JavaScriptType>]
type MainPage() =
    inherit Web.Control()
    [<JavaScript>]
    override this.Body =
        let result = P [Text ""]
        Div [
            P [Text "Press to retrieve data"]
            Input [Type "Button";
                   Value "Get Data"]
             |>! OnClick
                (fun e a -> result.Text <-
                            Application.GetPrimes())
            result
```

In this page we used Domain Specific Language (DSL) similar to HTML to declaratively layout the page. While this approach may not seem natural (using HTML itself, which is supported by layout tools, seems better option), it has the following advantages:

- DSL constructions can be grouped together to form higher-level layout abstractions, from which the page can be constructed
- We use this layout only for the inner part of the page, and the main design can be done in the traditional manner.

## B. Client Code Execution via Translation to Javascript

In the example above we attach client-side OnClick behavior to the button, which triggers the GetPrime() function. This function is declared in F# using special attributes that turn the function into the quotation:

```
module Application =
    [<JavaScript>]
    let rec primes =
        function
        | [] -> []
        | h::t -> h::primes(List.filter
                                (fun x -> x%h>0) t)
    [<JavaScript>]
    let GetPrimes() = primes [2..100]
```

Given [<JavaScript>] declaration WebSharper translates the function to JavaScript code inside the page, which looks similar to the following (here we give only part of the code corresponding to primes function):

```
(WebSharperProject$.Application$).primes =
function (_arg1)
{
  var __1;
  if (_arg1.$ == 1)
  {
    var t = _arg1.$1; var h = _arg1.$0;
    var l = t;
    var p = function (x) { return x % h > 0; };
    __1 =
new W$.$List({$: 1,$0: h, $1:
((WebSharperProject$.Application$).primes)(
((W$.Seq$).toList)(((W$.Seq$).filter)(p, l)))
                });}
  else { __1 = new W$.$List({ $: 0 }); }
  return __1; };
```

## C. Transparent Server-Side Code Execution

Another idea in WebSharper is to allow calling server-side functions transparently from the client code. In order to transfer the execution of primes function to the server side, we only need to modify the attributes of the function in the following manner:

```
module Application =
    let rec primes =
        function
        | [] -> []
        | h::t -> h::primes(List.filter
                                (fun x -> x%h>0) t)
    [<Rpc>]
    let GetPrimes() = primes [2..100]
```

This results in the different stub generated for primes function inside the JavaScript auto-generated code, which performs the actual remote call to the server. Please, note, that type translation and marshaling happens automatically and transparently, thus the end-developer does not need to worry about a lot of machinery behind the code distribution.

## D. Higher-Level Abstractions: Formlets, Flowlets, Pagelets

The fact that the page is put together in pieces by DSL constructs in one F# project allows for building higher-level functional abstractions over the basic principles outlined above. In particular, those abstractions allow us to solve many traditional tasks of web development, such as developing dynamic form-based user interface for handling data. Here is how we can create a very functional web form (called Formlet in WebSharper) for entering number with AJAX validation:

```
Div [
    formlet {
        let! max = Controls.Input "100"
            |> Validator.IsInt "Must be int"
        return max |> int
    }
    |> Enhance.WithTextLabel "Enter max number:"
    |> Enhance.WithValidationIcon
    |> Enhance.WithSubmitAndResetButtons
    |> Enhance.WithFormContainer
    |> Formlet.Run(fun n ->
            res.Text <- Application.GetPrimes(n))
    res ]
```

You can notice that formlet is described by a computational expression, another great meta-programming feature of F# [5,6]. In addition to formlets, there are also such abstractions as pagelets and flowlets for representing dialog flow between different pages.

## VI. DISTRIBUTED PROCESSING IN THE CLOUD

Finally, we want to emphasize that F# is also well suited for distributed cloud computing. Since functional style forces us to decompose the problem in such a way that it consists of stateless functions that compute the result given only the initial data, we can naturally run those independent functions in parallel on different nodes in the cluster. Distribution of the tasks is still the responsibility of the developer, but the whole nature of functional decomposition process makes it much easier.

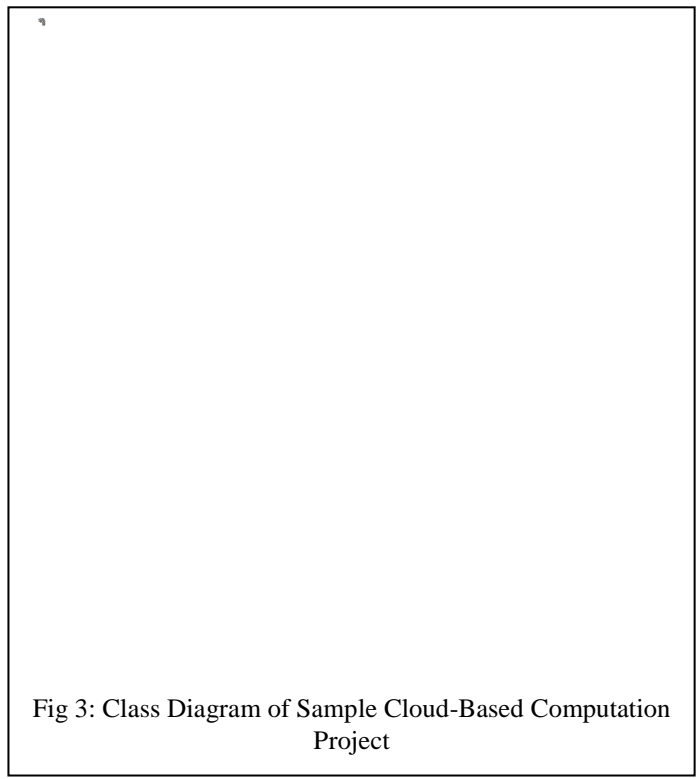

Fig 3: Class Diagram of Sample Cloud-Based Computation Project

To simplify the development of parallel computation applications for the cloud we have developed AzParCompute library (the main author of the library is Taymuraz Abaev, graduate student of Higher School of Economics) that can handle certain easily parallelizable scenarios, such as parametric sweep problems.

The library contains a template for Windows Azure computation worker role and a client library. In order to use the library, the developer needs to figure out the computation distribution mechanism and define types used to exchange the data between nodes, then to extend the base class for worker role and add the computation process itself. The library hides the complexity of communication between processes using Azure queues, so that the developer needs to concentrate on the problem.

To demonstrate the usage of cloud-based computation we have developed a sample application that uses Azure cloud to compute high-resolution Mandelbrot set and display it to the user. Whenever user changes resolution, the set is re-calculated at the new magnification by spawning several cloud-based computation instances. Each process computes a part of the set and returns corresponding partial bitmap, which are then combined on the client to form the resulting image. The client application is implemented in XNA, also using pure F#.

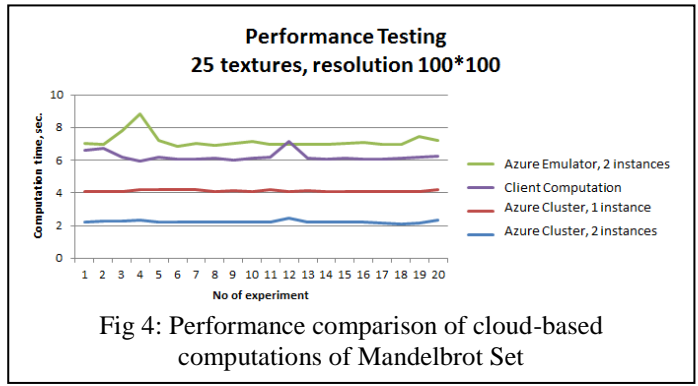

Fig 4: Performance comparison of cloud-based computations of Mandelbrot Set

Given the parametric sweep nature of the computation, we were able to get very good speedup by deploying computations on Azure. Doubling the number of instances decreased the computation time almost twice, using 100x100 pixels images.

It has to be noted, that using F# for Windows Azure Compute roles is supported by Visual Studio out of the box – so you do not need to use any specific libraries and tools in order to develop data-processing computational tasks in F#.

In addition to Windows Azure and the cloud, F# can be used to write parallel applications for computer clusters using MPI.NET. We will not cover this scenario in detail, referring our readers to [5], where a sample of cluster-based computation in F# is given.

## VII. CONCLUSION

In this paper, we have tried to show different successful examples of using functional programming and in particular F# programming language for the development of distributed applications on the web. While we do not have any quantitative research on the effectiveness of using functional programming in this problem domain, we believe that our readers can see the beauty of functional web development, and would consider using it in the real-life projects.

Many other advantages of using functional programming in real-life projects are outlined in [9], such as much reduced debugging time, more concise syntax, etc. Traditionally, one of the main disadvantages of functional approach was the performance, which used to be lower than that of imperative programs (mainly due to techniques such as garbage collection). However, given the fact that F# is the language of the .NET stack and it is translated to the same byte code, correctly programmed functional code in F# is almost as efficient, as C#. Thus, the only real drawback of using F# which would remain true for at least some time is the lack of qualified developers able to write and understand F# and functional paradigm. We hope that this paper would inspire more people to look at the functional programming, and thus would also indirectly help to bridge this gap.